\documentclass{aa}  

\usepackage[%
backref=true,%
pagebackref=true,%
hyperfootnotes=true,%
hyperindex=true,%
pageanchor=true,%
plainpages=false,%
colorlinks=true,%
linkcolor=blue,%
citecolor=blue,%
urlcolor=cyan,%
linktocpage=true,%
pdfstartview={Fit},
bookmarksopen=false
]{hyperref}
\usepackage{color}
\definecolor{pink}{rgb}{1,0,0.5}
\usepackage{hyperref}
\usepackage{graphicx, subfigure}

\usepackage{graphicx}
\usepackage{txfonts}

\begin{document} 

   \title{Spectral and morphological study of the gamma radiation of the middle-aged SNR HB 21}

   \author{L. Ambrogi
          \inst{1}
          \and
          R. Zanin
          \inst{1}
          \and 
          S. Casanova
          \inst{1, 2}
          \and
          E. De O\~na Wilhelmi
          \inst{3}
           \and
           G. Peron
          \inst{1}
          \and
          F. Aharonian
          \inst{1,4,5}
          }

   \institute{Max-Planck-Institut f\"ur Kernphysik, Saupfercheckweg 1, D-69117 Heidelberg, Germany\\
              \email{lucia.ambrogi@gmail.com}
           \and
             Institute of Nuclear Physics, Radzikowskiego 152, 31-342 Krakow, Poland
            \and 
            Institute for Space Sciences (CSIC/IEEC), E-08193 Barcelona, Spain
             \and
             Gran Sasso Science Institute, viale Francesco Crispi, 7 67100 L'Aquila, Italy
           \and
             Dublin Institute of Advanced Studies, 10 Burlington Road, Dublin 4, Ireland
             }

   \date{Received July 30, 2018; accepted January 10, 2019}

 
  \abstract
        {}
   {We investigate the nature of the accelerated particles responsible for the production of the gamma-ray emission observed from the middle-aged supernova remnant (SNR) HB 21.}
   {We present the analysis of more than nine years of \emph{Fermi} LAT data from the SNR HB 21. We performed morphological and spectral analysis of the SNR by means of a three-dimensional binned likelihood analysis. To assess the intrinsic properties of the parent particle models, we fit the obtained gamma-ray spectral energy distribution of the SNR by both hadronic- and leptonic-induced gamma-ray spectrum.}
   {We observe an extended emission positionally in agreement with the SNR HB 21. The bulk of this gamma-ray emission is detected from the remnant; photons up to $\sim$10\,GeV show clear evidence of curvature at the lower energies. The remnant is characterized by an extension of $0^{\circ}.83$, that is, 30\% smaller than claimed in previous studies. The increased statistics allowed us also to resolve a point-like source at the edge of the remnant, in proximity to a molecular cloud of the Cyg OB7 complex. In the southern part of the remnant, a hint of an additional gamma-ray excess in correspondence to shocked molecular clouds is observed.}
  {The spectral energy distribution of the SNR shows evidence of a break around 400 MeV, which can be properly fitted within both the hadronic and leptonic scenario. The pion-decay mechanism reproduces well the gamma rays, postulating a proton spectrum with a slope $\sim 2.5$ and with a steepening around tens of GeV, which could be explained by the energy-dependent escape of particles from the remnant.  In the leptonic scenario the electron spectrum within the SNR matches closely the locally measured spectrum. This remarkable and novel result shows that SNR HB 21 could be a direct contributor to the population of Galactic electrons. In the leptonic scenario, we find that the local electron spectrum with a break around 2 GeV, closely evokes the best-fitting parental spectrum within this SNR. If such a scenario is confirmed, this would indicate that the SNR might be a source of Galactic background electrons.}

\maketitle

\section{Introduction}
\label{sec:Intro}

Supernova remnants (SNRs) are believed to be sources of Galactic cosmic rays (CRs). According to the SNR paradigm, CRs are accelerated through diffusive shock acceleration (DSA) at SNR shocks and propagate diffusively in the Galaxy \citep{bell1978, blandford78}. A confirmation of this acceleration scenario can be provided by gamma-ray observations: accelerated CRs might undergo interactions with the matter field in the region of the source, leading to the production of hadronic gamma-ray photons through $\pi^0$-decay or leptonic gamma-ray photons through bremsstrahlung processes. Relativistic CR electrons might also produce leptonic gamma rays through inverse Compton scattering off the photon field. The spectral energy distribution (SED) of the hadronic gamma rays is expected to show a peculiar feature, the so-called pion bump, which is an abrupt turnover of the spectrum that rises steeply below 200\,MeV due to the finite rest mass of the neutral pions. At the time of writing, this spectral feature is claimed to be detected in three bright SNRs: IC 443 \citep{Ackermann:2013wqa}, W44 \citep{Giuliani:2011nx, Ackermann:2013wqa}, and W51C \citep{Jogler:2016lav}. All of these three objects are middle-aged SNRs interacting with molecular clouds, characterized by a shell-like radio morphology combined with a center-filled X-ray morphology. 

Other middle-aged SNRs have been successfully detected so far. The SNR HB 21 (also known as G89.0+4.7) belongs to this class of objects and is a few tens of thousands years in age \citep{koo1991survey, leahy1996rosat}. Similar to IC 443, W44, and W51C, HB 21 is also a mixed morphology SNR. Its radio shell has an angular size of $120' \times 90'$, slightly tilted in the northwest-southeast direction \citep{hill1974observations}. Thermal X-ray emission is concentrated in the interior of the radio shell and appears elongated in the same direction \citep{Lazendic:2005tr, Pannuti:2010fg}. Most previous studies assumed 0.8\,kpc as the distance to HB 21, postulating an association with the molecular clouds in the Cyg OB7 complex on the eastern side of the SNR \citep{tatematsu1990, koo2001}. Indeed, on the eastern boundary of the SNR, $\mbox{HI}$ observations revealed a rather uniformly distributed emission (the so-called wall) as well as three large molecular clouds (clouds A, B and C hereafter) detected through CO observations \citep{tatematsu1990}. However, no direct kinematic evidence of interaction between the eastern clouds and the SNR shell was observed. By comparing the $\mbox{HI}$ and CO column density with the X-ray absorbing column density, the distance was later estimated to be $\sim$1.7\,kpc  \citep{byun2006}. We take this as the distance to HB 21 throughout this paper. Broad line emissions have been detected from shocked molecular clumps in the northern and southern regions of the SNR (clouds N and S) \citep{koo2001}, as well as in the northwest (cloud NW) \citep{byun2006}, providing the most solid evidence for the SNR interaction with molecular clouds. The interaction with northern and southern clouds is confirmed by observations in other wavelengths: the lack of thermal X-ray emission in correspondence to these shocked molecular clumps suggests that the shock of the SNR could have gone radiative owing to interactions with molecular clouds \citep{Lazendic:2005tr}. A recent study based on X-ray observations of the inner part of the SNR interpreted the existence of recombining thermal plasma as due to interactions with molecular clouds, which causes weaker magnetic turbulence, permits the escape of CR protons, and cools down the thermal electron \citep{suzuki2018discovery}. Moreover, near- and mid-infrared images of the same regions showed shock-cloud interaction features \citep{shinn2009infrared, Shinn:2009zn}. 

In the second \emph{Fermi} LAT source catalog, three point-like sources were initially associated with the SNR, interpreted as local maxima of the extended G89.0+4.7 region; these sources are 2FGL J2041.5+5003, 2FGL J2043.3+5105, and 2FGL J2046.0+4954 \citep{Fermi-LAT:2011yjw}. Later studies, also considered a fourth point-like object lying very close to the northeast edge of the SNR shell (2FGL J2051.8+5054) as part of the SNR and determined that the gamma-ray emission is produced by a single extended source \citep{Reichardt:2012nx, Pivato:2013uoa}. Consequently, the SNR has been included in the third \emph{Fermi} LAT catalog (labeled as 3FGL J2045.2+5026e) as one of the 12 SNRs firmly identified as spatially extended objects \citep{3FGL}. According to roughly four years of Pass-7 data, its extended emission is well modeled as uniform disk with radius $1^{\circ}.19$ and centered at $l=88^{\circ}.75, b=+4^{\circ}.65$ and has a SED best described by a log parabola \citep{Pivato:2013uoa}. 

In this paper we present an analysis based on more than nine years of the public \emph{Fermi} LAT data from SNR HB 21 and we discuss the nature of its emission mechanism. In Section \ref{sec:analysis} we describe the data used for the study and their preparation. In Section \ref{sec:results} we report the morphological and spectrometric analysis performed on the data. In Section \ref{sec:discussion} we discuss the results and their interpretation. Finally, conclusions are summarized in Section \ref{sec:conclusions}.

\section{Data analysis}
\label{sec:analysis}

The analysis presented is based on Pass-8 data collected from the beginning of the science operations of the instrument, on August 4 2008, to October 13 2017, corresponding to $\sim9.2$ years of data. The analysis of the data sample was performed using Fermipy \citep{Wood:2017yyb}, an open-source python package built on the \emph{Fermi Science Tools,} which provides a high-level interface for analyzing LAT data. The LAT \emph{Science Tools} exploited in this work is version \texttt{v10r0p5}, available at the \emph{Fermi} Science Support Center\footnote{\url{https://fermi.gsfc.nasa.gov/ssc/}}. We selected class 128 events with  reconstructed energy from 60\,MeV to 870\,GeV.
We did not perform any selection in terms of point spread function (PSF) quality and energy reconstruction quality and we considered both front and back photon-to-pair conversion layer (\texttt{evtype=3}). We selected good time intervals requiring the instrument configuration to be in the science mode (\texttt{LAT\_CONFIG==1}) and removing data with bad quality flag (\texttt{DATA\_QUAL>0}). Moreover, to filter out photons coming from the Earth's limb, we rejected events with a zenith angle larger than 90$^{\circ}$. We considered a region of interest (ROI) defined by a circle centered at the position of 3FGL J2045.2+5026e and with a radius of 10$^{\circ}$. To compensate for the large PSF of the detector at the low energies, we account for the exposure of nearby objects by considering the contribution of gamma-ray emitters up to 20$^{\circ}$ from the ROI center. 

\begin{figure}[htb]
\centering
\includegraphics[width=0.5\textwidth]{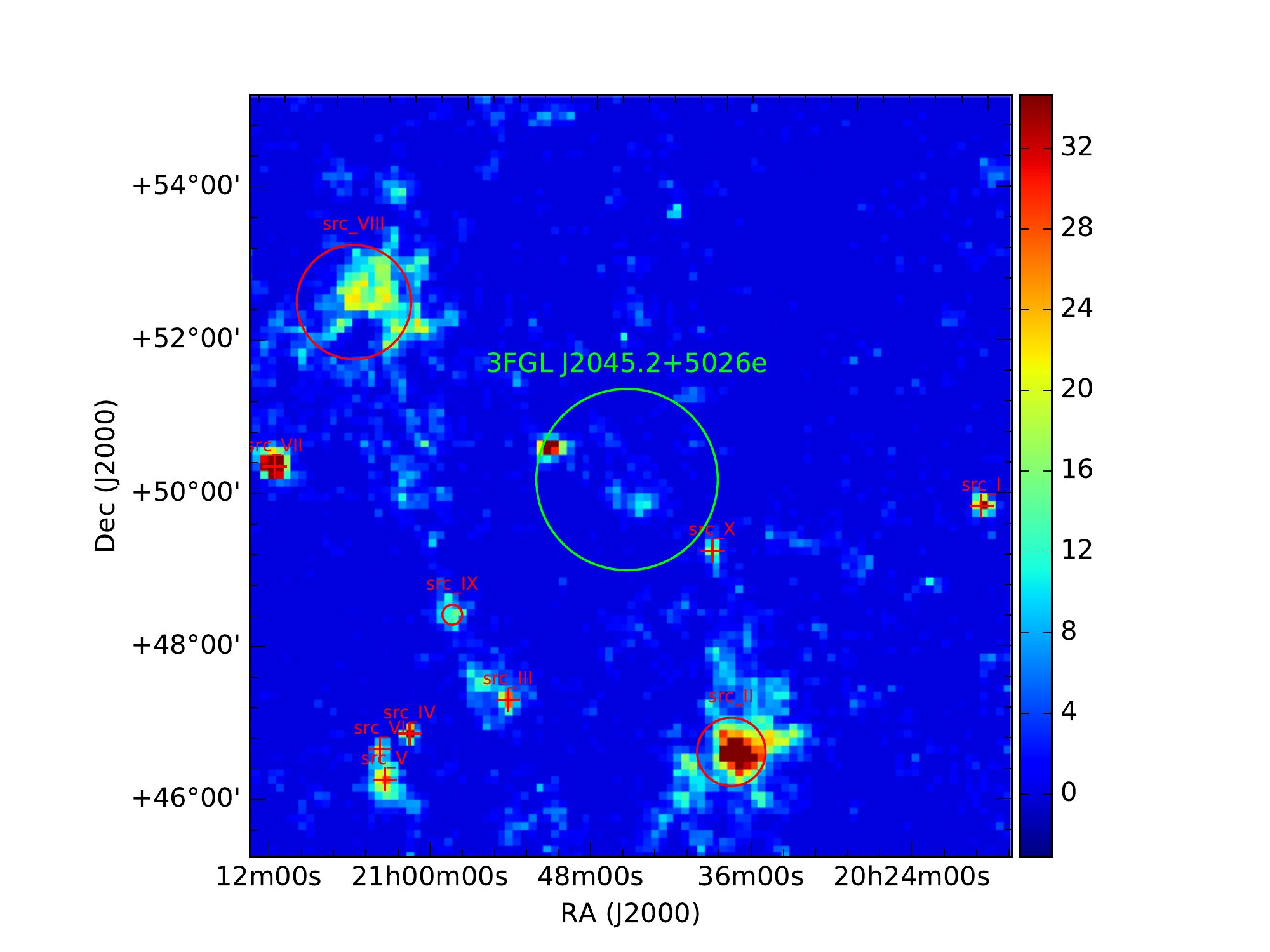}
\caption{\label{fig:newSrc}Test statistic map of the selected ROI with the new gamma-ray excesses found and modeled within this analysis. Red crosses indicate point-like sources, while red circles quantify the estimated extensions of the excesses (see Table \ref{tab:newSrc}). The position and the extension of the SNR HB 21 as defined in the 3FGL catalog is shown in green. }
\end{figure}

We study the morphology and the spectral properties of HB 21 by means of a three-dimensional binned likelihood analysis: for the two spatial dimensions, the data are binned using square bins of $0^{\circ}.1$ side, while for the third dimension we assume ten logarithmically-spaced energy bins per decade. We verified the validity of the morphological results using PSF event types in a joint likelihood analysis. The initial model includes the current Galactic diffuse emission model (\texttt{gll\_iem\_v06.fits}) and the corresponding model for the Extragalactic isotropic diffuse emission (\texttt{iso\_P8R2\_SOURCE\_V6\_v06.txt}), both of which are provided in the \emph{Science Tools}\footnote{\url{https://fermi.gsfc.nasa.gov/ssc/data/access/lat/BackgroundModels.html}}, as well as the gamma-ray sources in the 3FGL catalog. We left the normalization of the Galactic and extragalactic diffuse models as a free parameter. The same was also done for 3FGL objects located within 5$^{\circ}$ from the source center, while for those at distances smaller than 3$^{\circ}$ we free all the spectral parameters. The resulting test statistic (TS) map reveals a number of significant excesses in the ROI, as shown in Fig. \ref{fig:newSrc}. We localized these gamma-ray emissions, which are not included in the 3FGL model, and estimated their extension (see Table \ref{tab:newSrc}). In a second iteration step the new hotspots have been added to the background model. We removed 3FGL J2045.2+5026e from such a model and we refer to it as the null hypothesis, as shown in Fig. \ref{fig:newSrc}.
\begin {table*}[h!]
\caption { \label{tab:newSrc} List of new gamma-ray excesses found in the ROI of the analysis. In the last column, the name of a possibly associated source is reported, whether a spacial agreement with the newly detected gamma-ray excess is found.}
\begin{center}
\begin{tabular}{l  c  c  c}
\hline
 Excess & $(l,b)$ & Extension & Association name\\
\hline  
 \texttt{src\_I} & \scriptsize{$85^{\circ}.57\pm0^{\circ}.02, +8^{\circ}.10\pm0^{\circ}.02$} & -  & \scriptsize{1RXS J201619.3+495323}\\
 \texttt{src\_II} & \scriptsize{$85^{\circ}.09\pm0^{\circ}.02, +3^{\circ}.52\pm0^{\circ}.03$} & \scriptsize{$0^{\circ}.29^{+0^{\circ}.16}_{-0^{\circ}.07}$} & \scriptsize{MITG J2036+4654}  \\ 
 \texttt{src\_III} & \scriptsize{$87^{\circ}.45\pm0^{\circ}.04, +1^{\circ}.65\pm0^{\circ}.03$}  & -  & \scriptsize{NVSS J205432+473338}\\
 \texttt{src\_IV} & \scriptsize{$87^{\circ}.90\pm0^{\circ}.02, +0^{\circ}.36\pm0^{\circ}.02$}  & - & \scriptsize{MITG J2102+4702}\\
 \texttt{src\_V} & \scriptsize{$87^{\circ}.63\pm0^{\circ}.04, -0^{\circ}.27\pm0^{\circ}.04$}  & -  & - \\
 \texttt{src\_VI} & \scriptsize{$87^{\circ}.98\pm0^{\circ}.09, -0^{\circ}.07\pm0^{\circ}.10$}  & -  & - \\
 \texttt{src\_VII} & \scriptsize{$91^{\circ}.75\pm0^{\circ}.02, +1^{\circ}.14\pm0^{\circ}.02$}  & -  & \scriptsize{29P 68} \\
 \texttt{src\_VIII} & \scriptsize{$92^{\circ}.81\pm0^{\circ}.08, +3^{\circ}.29\pm0^{\circ}.08$}  & \scriptsize{$0^{\circ}.75\pm0^{\circ}.04$} & -  \\
 \texttt{src\_IX} & \scriptsize{$88^{\circ}.78\pm0^{\circ}.12, +1^{\circ}.77\pm0^{\circ}.07$}  & \scriptsize{$0^{\circ}.13\pm0^{\circ}.05$} & - \\
 \texttt{src\_X} & \scriptsize{$87^{\circ}.32\pm0^{\circ}.08, +4^{\circ}.97\pm0^{\circ}.07$}  & -  & -\\
\hline
\end{tabular}
\end{center}
\end{table*}

\section{Results}
\label{sec:results}

\subsection{Morphology}
\label{sec:morpho}

\begin{figure*}[htb!]
\centering
\subfigure[]{\label{fig:morpho_full}\includegraphics[width=0.49\textwidth]{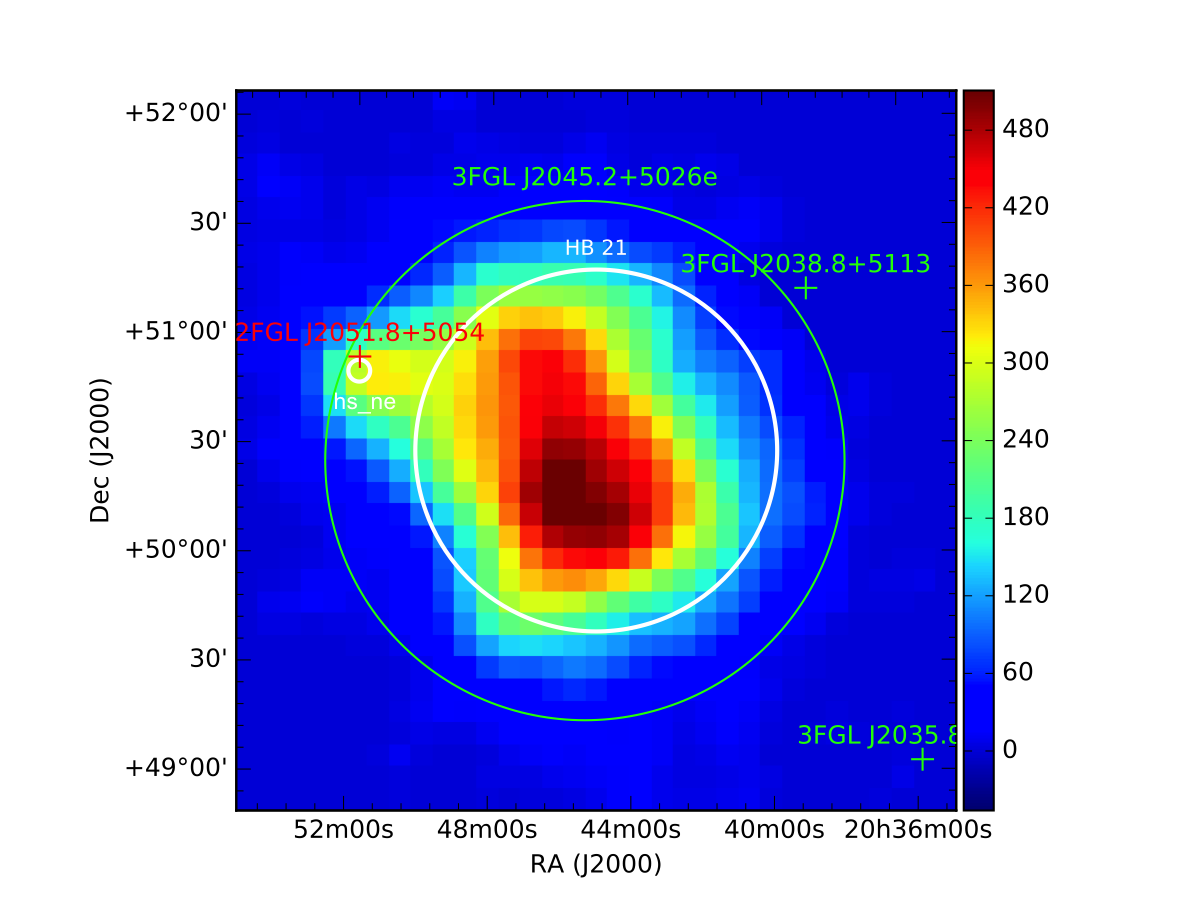}}
\subfigure[]{\label{fig:morpho_500}\includegraphics[width=0.49\textwidth]{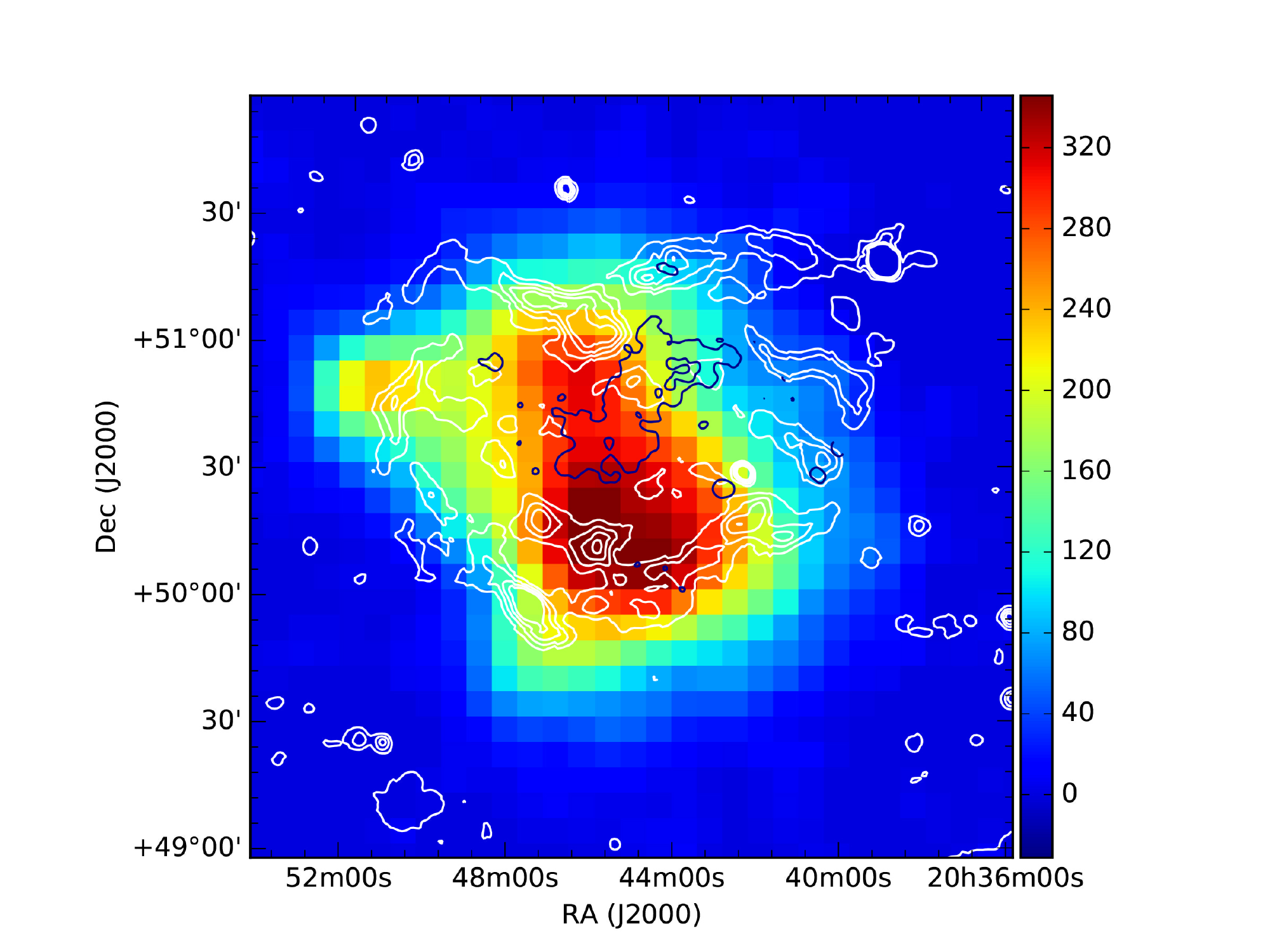}} \\
\subfigure[]{\label{fig:morpho_1000}\includegraphics[width=0.49\textwidth]{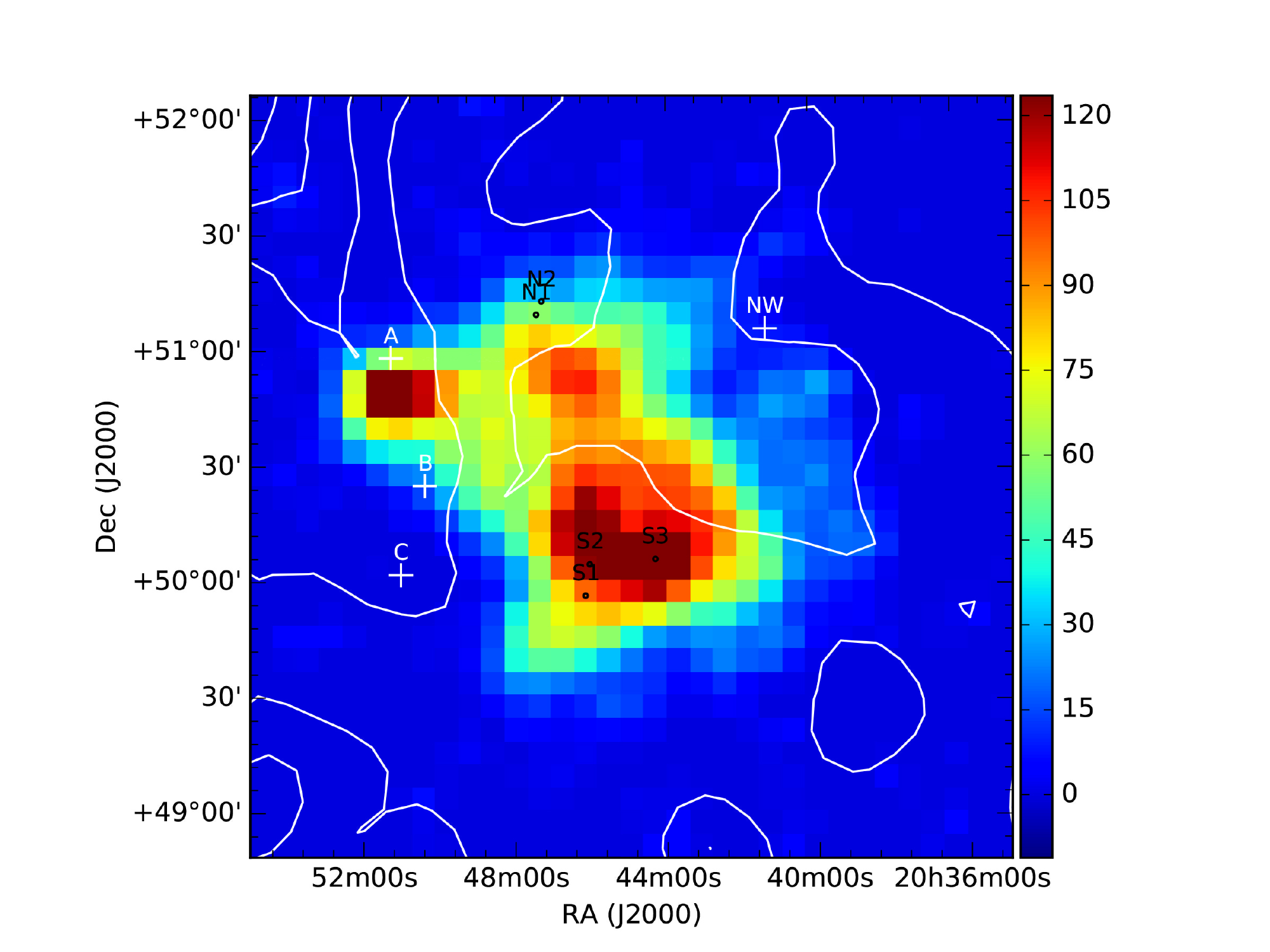}} 
\subfigure[]{\label{fig:morpho_residual}\includegraphics[width=0.49\textwidth]{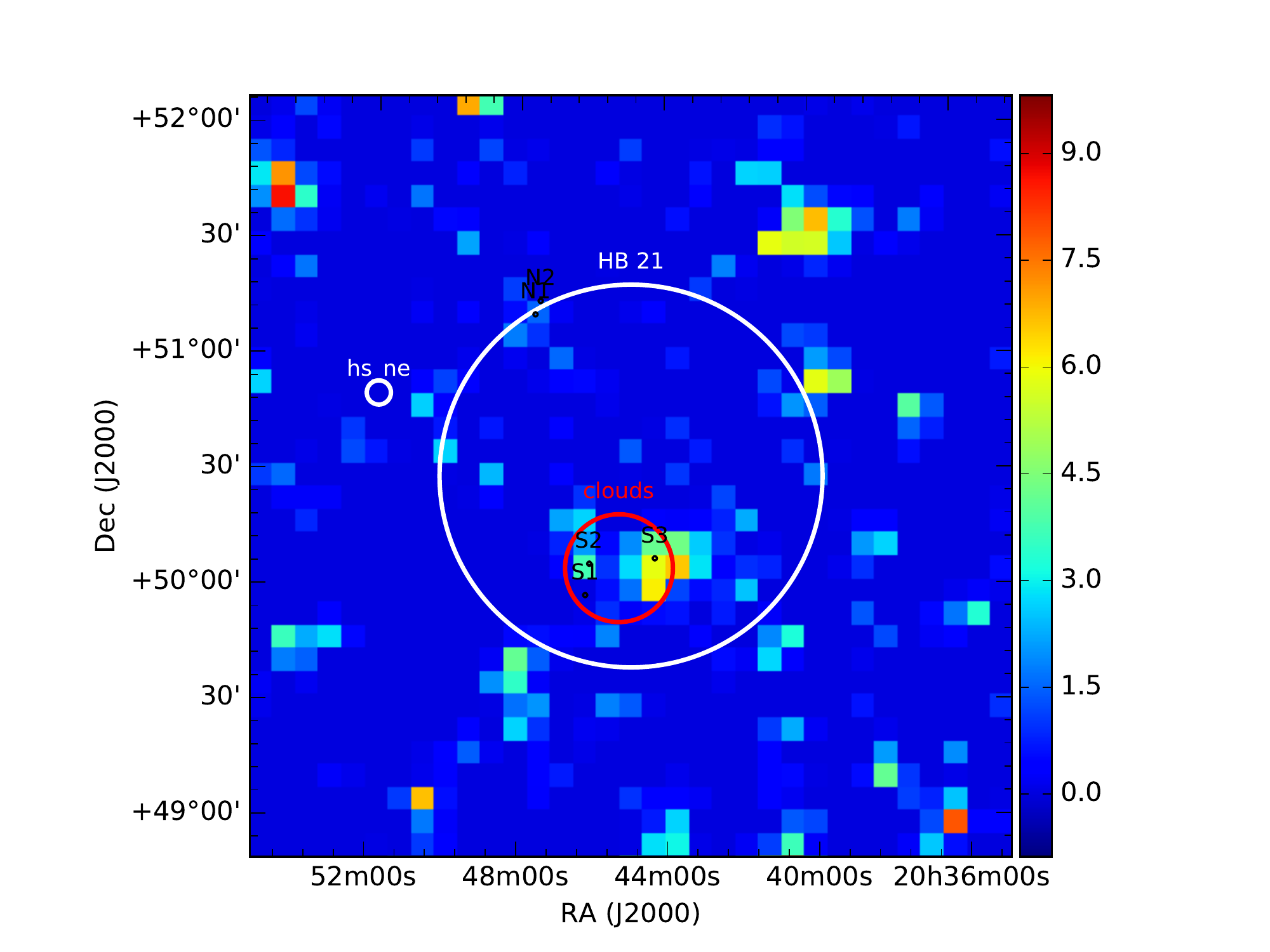}}
\caption{Test statistic map of the selected ROI around the SNR HB 21. The map in Fig. \ref{fig:morpho_full} is computed in the whole energy interval considered for the analysis, i.e., above 60\,MeV. Overlaid to this map, the 3FLG sources are reported;  3FGL J2045.2+5026e is also included and defined in the catalog as a disk-type geometry with a radius of $1^{\circ}.19$ (green markers). The best-fit model for the SNR HB 21 as found in this work is also shown (disk-type geometry with a radius of $0^{\circ}.83$), together with the nearby hotspot \emph{hs\_ne} (white regions). The position of the point source 2FGL J2051.8+5054 is presented as well (red cross). In Fig. \ref{fig:morpho_500} the TS map above 500\,MeV is presented,  together with the contours of radio continuum emission at 4850 MHz (white) and the ROSAT X-ray (dark blue) contours. The ROSAT data are available at \url{https://skyview.gsfc.nasa.gov/current/cgi/titlepage.pl}. The map in Fig. \ref{fig:morpho_1000} is for energies above 1\,GeV. On top of the TS map, the CO contours are reported in white. The CO distribution are taken from \cite{Dame:2000sp} and integrated between $-20\mbox{ km s}^{-1}$ and $10\mbox{ km s}^{-1}$: six contour levels are linearly spaced from $1.5\mbox{ K km s}^{-1}$ to $28\mbox{ K km s}^{-1}$. We also show the position of the shocked CO clumps S1, S2, S3, N1, N2 (black circles) \citep{koo2001} and clouds A, B, C \citep{tatematsu1990} and NW \citep{byun2006} (white crosses). In Fig. \ref{fig:morpho_residual} the TS map of the residual emission of the best modeling of the SNR and the hotspot (white regions) is shown above 1\,GeV. In addition, the position of the shocked CO clumps is reported, together with the estimated cloud region (red region).}
\label{fig:morpho}
\end{figure*}

Several scenarios describing the morphology of HB 21 were considered and tested as follows:
\begin{itemize}
\item[1.] a disk-type geometry with fixed radius of $1^{\circ}.19$, as estimated in Ref. \citep{Pivato:2013uoa};
\item[2.] a disk-type geometry with free radius;
\item[3.] a radial Gaussian shape with free width;
\item[4.] a ring-type geometry defined on the basis of the radio contours of the SNR at 4850 MHz (see Fig. \ref{fig:morpho_500}), i.e., an elliptical ring  $40^{\circ}$ tilted in the northwest-southeast direction with inner radii of $0^{\circ}.3$ and $0^{\circ}.45$ and outer radii of $0^{\circ}.9$ and $1^{\circ}.05$.
\end{itemize}
The morphology of the source was studied under a certain assumption on the shape of its spectral profile. We consider a log parabola energy distribution, which corresponds to the best description of the spectrum reported in the catalog \citep{3FGL, Pivato:2013uoa}. The disk-type morphology, with the radius as free parameter, is the preferred model and has a significance of $44\sigma$ with respect to the null hypothesis.

This model predicts a radius smaller than that previously published \citep{Pivato:2013uoa}. Indeed, the TS map revealed the presence of a hotspot very close to HB 21, placed northeast with respect to HB 21 center and hereafter called \emph{hs\_ne}.
The hotspot is well described as a point-like emitter located at $l=89^{\circ}.72\pm 0^{\circ}.02, b=+4^{\circ}.11\pm0^{\circ}.02$ and with a power-law spectrum $dN/dE = N_0 (E/E_0)^{-\alpha}$, with  $E_0 = 1000$\,MeV and estimated normalization $N_0= (0.18 \pm 0.02) \cdot 10^{-12}$ MeV$^{-1}$cm$^{-2}$s$^{-1}$ and photon index $\alpha =2.28\pm0.07$. The position of \emph{hs\_ne} is in good agreement with the gamma-ray source 2FGL J2051.8+5054 ($l=89^{\circ}.77, b=4^{\circ}.15$), as shown in Fig. \ref{fig:morpho_full}. This 2FGL source was removed in the following catalogs because it was included as part of the extended SNR HB 21 \citep{Reichardt:2012nx,Pivato:2013uoa}. Conversely, we found that the significance of the hypothesis of \emph{hs\_ne} to be separated from the SNR is $\sim$9$\sigma$. Once the hotspot is included in the model as an independent source, the extension of HB 21 is eventually estimated to be $0^{\circ}.83 \pm 0^{\circ}.02$ and centered at $l=88^{\circ}.76\pm 0^{\circ}.02, b=+4^{\circ}.74\pm0^{\circ}.02$, as shown in Fig. \ref{fig:morpho_full}.

We further investigate the morphology of the source by testing the hypothesis of two extended objects, one located in the north and one in the south with respect to the center of the source. This configuration is disfavored at $3\sigma$ level with respect to the single disk-type morphology, providing no indication for the existence of multiple sources within the region of SNR HB 21. In Fig. \ref{fig:morpho_residual}, the residuals of the modeling with the single disk-type geometry are shown. An excess of gamma-ray emission is left in the southern part of the source, interestingly close to the region of the shocked CO clumps \citep{koo2001}. To assess the significance of this excess, we postulate the existence of an additional component placed on top of the whole disk and located at $l=88^{\circ}.47, b=4^{\circ}.45$, which is the mean position of the three clumps S1, S2, S3. Its extension is estimated to be $0^{\circ}.24 \pm 0^{\circ}.05$ (red region in Fig. \ref{fig:morpho_residual}) and has a power-law spectrum with parameters $E_0 = 1000$\,MeV, $N_0= ( 0.76 \pm 0.21) \cdot 10^{-12}$ MeV$^{-1}$cm$^{-2}$s$^{-1}$ and $\alpha = 2.32 \pm 0.12$. Only marginal evidence ($4\sigma$) of additional emission is found in correspondence to this cloud region, which does not allow us to establish firmly the presence of this additional component. These results, together with the fact that the gamma-ray source is produced by a single extended source, suggest that the bulk of the detected gamma-ray emission originates from the remnant and the contribution from the shocked clouds can be neglected.

\subsection{Spectral analysis}
\label{sec:spec}

\begin{figure}[htb]
\centering
\includegraphics[width=0.5\textwidth]{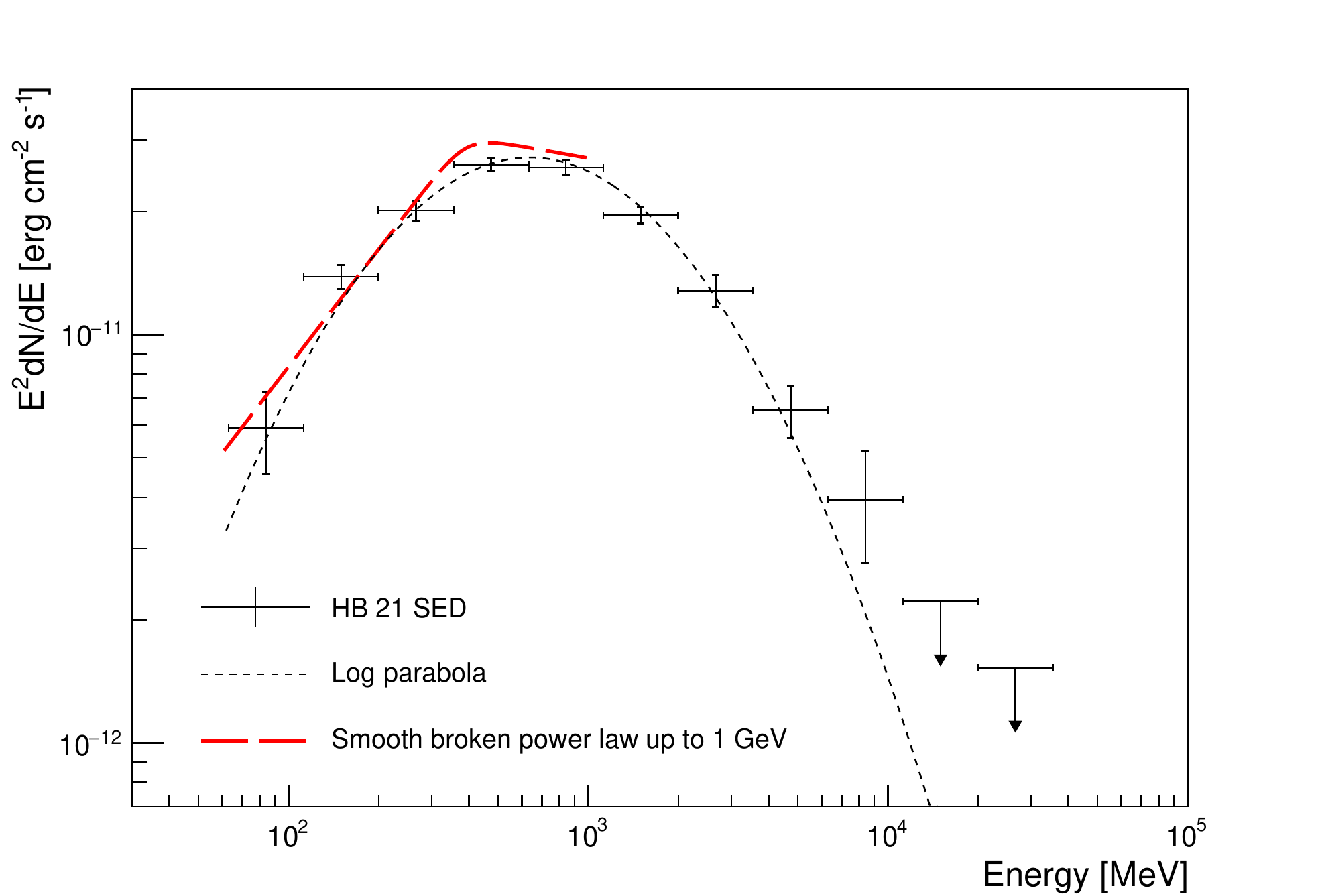}
\caption{\label{fig:SED_HB21}SNR HB 21 SED. The black dashed curve indicates the global best-fit model, i.e., the log parabola spectral shape. The red dashed curve up to 1 GeV is the broken power-law fit.}
\label{fig:new_src}
\end{figure}

We compute the SED of the source considering four independent logarithmic bins per decade of energy. In each energy bin, the spectral points are obtained performing a fit according to a power-law profile with spectral index $\alpha=2$. The corresponding SED of the SNR is shown in Fig. \ref{fig:SED_HB21}.

Our spectral analysis confirms that the spectrum shows clear evidence of curvature, as previously found in \cite{Reichardt:2012nx,Pivato:2013uoa}, and provides for the first time a solid estimation of the statistical significance of the falloff. To account for the curved gamma-ray emission over the whole energy domain, we tested several parameterizations of the spectrum in this analysis: (i) a log parabola; (ii) a smooth broken power law; and (iii) a power law with an exponential cutoff. In Table \ref{tab:ts_spectra} the results of the fit are reported. Since the spectral models considered are not nested, we use the Akaike information criterion as an estimator of the relative quality of the models: $\mbox{AIC} = 2k - 2\log\mathcal{L}$, with $k$ the number of estimated parameters in the model and $\log\mathcal{L}$ the maximum value of the likelihood function \citep{akaike1974new}. The favored model, which corresponds to the the minimum AIC value, is the log parabola, defined as $dN/dE = N_0 (E/E_0)^{- (\alpha +\beta\ln(E/E_0))}$, where $E_0 = 1000$\,MeV and estimated parameters $\alpha= 2.35\pm0.03$, $\beta=0.39\pm0.03$ and $N_0= (0.157 \pm 0.004) \cdot 10^{-10}$\,MeV$^{-1}$cm$^{-2}$s$^{-1}$. The analytical description of the spectrum according to the log parabola function makes it possible to reproduce locally the observed steepening of the gamma-ray energy distribution, which moves from $\alpha\sim0.5$ between 60 and 200\,MeV, to $\alpha\sim1$ from 200 to 400\,MeV, then to $\alpha\sim2$ from 400 and 1000\,MeV and then to $\alpha\sim3$ above 1000\,MeV. Globally, with respect to the catalog values \citep{Pivato:2013uoa}, our results show a spectrum with a slightly harder slope and weaker flux; this difference is explained by the different extraction regions used for the estimation of the gamma-ray emission from the SNR.
\begin {table}[h!]
\caption {Values of estimated spectral parameters of the global fit, with the parameter $N_0$ is given in units of MeV$^{-1}$cm$^{-2}$s$^{-1}$. Last column shows the AIC number of the model.} \label{tab:ts_spectra} 
\begin{center}
\begin{tabular}{l  c  c }
\hline
 Spectral model & Spectral parameters & AIC \\
\hline  
{\scriptsize Log parabola} & {\scriptsize $N_0= (0.157 \pm 0.004) \cdot 10^{-10}$}  &  {\scriptsize $-4323878$} \\
 & {\scriptsize $\alpha= 2.35\pm0.03$} & \\
 & {\scriptsize $\beta=0.39\pm0.03$} & \\
 {\scriptsize Smooth broken}  & {\scriptsize $N_0= (6.78 \pm 0.97) \cdot 10^{-10}$} & {\scriptsize $-4323862$} \\
 {\scriptsize power law} &  {\scriptsize $\alpha_1=  1.43 \pm 0.10$}  &\\
 &  {\scriptsize $\alpha_2= 2.90 \pm 0.08$} & \\
 &   {\scriptsize $E_b= 750 \pm 75$\,MeV} & \\ 
{\scriptsize Power law with}  &  {\scriptsize $N_0= (3.34 \pm 0.18) \cdot 10^{-10}$ } & {\scriptsize $-4323842$} \\
 {\scriptsize exponential cutoff}   & {\scriptsize $\alpha= 1.39 \pm 0.08$} & \\
 & {\scriptsize $E_c= 1.26\pm 0.13$\,GeV} &\\ 
 \hline
\end{tabular}
\end{center}
\end{table}

In order to assess the significance of the observed low energy falloff of the spectrum, we perform a likelihood fit of HB 21 in the energy range between 60\,MeV and 1\,GeV assuming: (i) a simple power-law in the form $dN/dE = N_0 (E/E_0)^{-\alpha}$, where $E_0 = 200$\,MeV; and (ii) a smooth broken power law parametrized as $dN/dE = N_0 (E/E_0)^{-\alpha_1} \left( 1 + (E/E_b)^{(\alpha_1 - \alpha_2)/\beta} \right)^{-\beta}$, where $E_0 = 200$\,MeV and $\beta=0.1$. The improvement in the likelihood fit achieved with the smooth broken power-law profile is estimated through the TS of the two nested models. We prove that the break is detected with a significance of $5.4\sigma$ and its characteristic energy is estimated to be $E_b = 389\pm95$\, MeV, with a change in photon index from $\alpha_1 = 1.04\pm0.28 $ to $\alpha_2 = 2.13\pm0.20$, and a normalization $N_0 = (2.53 \pm 0.25) \cdot 10^{-10}$\,MeV$^{-1}$cm$^{-2}$s$^{-1}$. 

\section{Discussion}
\label{sec:discussion}

\subsection{Modeling of the nonthermal emission}
\label{sec:modeling}

The gamma-ray emission from remnants can originate from different mechanisms, induced either by proton interactions or by electron interactions. To investigate the nature of the parent particles, we fit the obtained gamma-ray SED of the SNR by both hadronic- and leptonic-induced gamma-ray spectrum.  We use the Naima package (version \texttt{v0.8.1}) to compute the nonthermal radiation \citep{naima}. The gamma-ray emission resulting from pion-decay and bremsstrahlung processes are shown In Fig. \ref{fig:hb21_models}.  Different particle parent spectra are considered for the two scenarios.  The Bayesian information criterion (BIC) enables us to select a model on a statistic basis; it is defined as $\mbox{BIC} = \log(n)k - 2\log\mathcal{L}$, where $n$ is the number of data points, $k$ the number of parameters estimated by the model, and $\log\mathcal{L}$ the maximum value of the likelihood function \citep{Schwarz:1978tpv}. The model with the lowest BIC is preferred. In Table \ref{tab:naima_results_pion} and  \ref{tab:naima_results_brem} the BIC values are shown together with the estimated parameters of the parent particle spectra for pion-decay and bremsstrahlung mechanism, respectively. In the following we report a discussion for both the two emission processes.
\begin{figure*}[htb!]
\centering
\subfigure[]{\label{fig:pion}\includegraphics[width=0.49\textwidth]{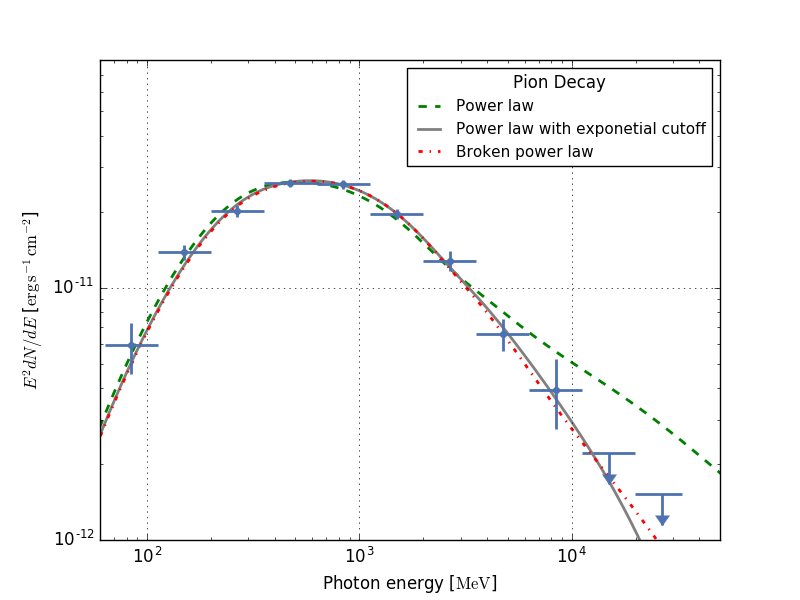}}
\subfigure[]{\label{fig:brem}\includegraphics[width=0.49\textwidth]{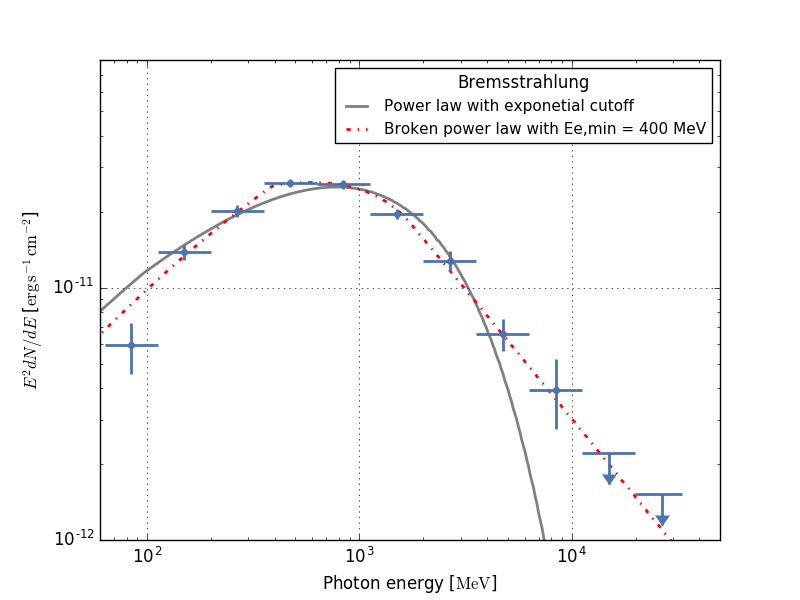}}
\caption{\label{fig:hb21_models} Emission modeling of the gamma-ray radiation from HB 21. In Fig. \ref{fig:pion}, the pion-decay process is shown, while in Fig. \ref{fig:brem} the bremsstrahlung mechanism is reported.}
\label{fig:hb21_models}
\end{figure*}

\begin {table*}[h!]
\caption {Fit results of the pion-decay production mechanism.} 
\label{tab:naima_results_pion} 
\begin{center}
\begin{tabular}{l c  c  c c c}
\hline
Spectral model & $\alpha_{1,p}$ & $\alpha_{2,p}$ & $E_{c,p} $\,(GeV)  & $E_{b,p} $\,(GeV)  &  BIC \\
\hline  
\scriptsize{Power law} & \scriptsize{$2.78 \pm 0.03$} &  - & -  & - &  \scriptsize{25.4}  \\
\scriptsize{Exponential cutoff} &  \scriptsize{$2.49^{+0.11}_{-0.14}$} & - & \scriptsize{$51^{+25}_{-20}$} &   -  &  \scriptsize{12.3} \\
\scriptsize{Broken power law} & \scriptsize{$2.53^{+0.11}_{-0.20}$} &  \scriptsize{$3.37^{+0.32}_{-0.20}$} & - & \scriptsize{$15^{+10}_{-7}$} &  \scriptsize{14.5} \\
\hline
\end{tabular}
\end{center}
\end{table*}
\begin {table*}[h!]
\caption {Fit results of the bremsstrahlung production mechanism.} 
\label{tab:naima_results_brem} 
\begin{center}
\begin{tabular}{l  c  c  c c c c }
\hline
Spectral model & $\alpha_{1,e}$ & $\alpha_{2,e}$ & $E_{c,e} $\,(GeV)  & $E_{b,e} $\,(GeV)  &  BIC \\
\hline  
\scriptsize{Exponential cutoff} &  \scriptsize{$0.79^{+0.14}_{-0.17}$} & - & \scriptsize{$1.6\pm0.2$} &  -  &  \scriptsize{45.2}\\
\scriptsize{Broken power law with $\mbox{E}_{e,min}= 400$\,MeV} & \scriptsize{$1.6 \pm 0.3$} &  \scriptsize{$3.14^{+0.17}_{-0.06} $} & -  & \scriptsize{$1.8^{+0.5}_{-0.3}$}  & \scriptsize{15.4}  \\
\hline
\end{tabular}
\end{center}
\end{table*}

\subsubsection{Pion decay}
The hadronic models shown in Fig. \ref{fig:pion} provide a good agreement with the gamma-ray data. In case of proton-proton interactions, the low energy break of the photon spectrum comes naturally from the kinematics of the process. However, a simple power-law parent spectrum fits the data only marginally, as it does not describe properly the high energy part of the spectrum. Consequently, a high energy steepening in the protons spectrum has to be introduced, either in the form of a cutoff or in the form of a break. The available statistics does not allow us to discriminate between a power law with an exponential cutoff or a broken power law. Under the hypothesis of hadronic-dominated scenario, this underlying proton spectrum has a spectral index on the order of $\alpha_{1,p} \sim 2.5$. Similar properties are found also in the SNRs IC 443, W44 \citep{Ackermann:2013wqa}, and W51C \citep{Jogler:2016lav}, for which a proton index steeper than $\alpha_{p} = 2$ is established. Compared to the other three SNRs, the proton and the photon indexes are slightly steeper in HB 21, while the energy of the break in the gamma-ray spectrum is higher; this is still within statistical uncertainties however. The steepening of the spectrum can be explained in terms of confinement time of particles in the source, as the escape of the most energetic particles might lead to a quasi-stationary spectrum that is significantly softer than the acceleration spectrum. Therefore, depending on the age of the accelerator and on its distance and diffusion coefficient, different gamma-ray spectra are expected from source to source \citep{aharonian1996emissivity}. 

In case of hadronic-dominated models, the CRs released by the SNR can escape the remnant and interact with the surrounding gas environment. Consequently, molecular clouds in the vicinity of the SNR can be illuminated by such escaping CRs and an enhanced gamma-ray emission is expected as due to the increased rate of $\pi^0$ decays \citep{Gabici:2009ak}. In this scenario, the gamma-ray brightness of hotspot \emph{hs\_ne} might be explained in terms of CRs escaping from HB 21 and interacting with the large molecular clouds along the eastern boundary of the remnant, which seem to be associated with the Cyg OB7 complex \citep{tatematsu1990}. In particular, the northern cloud A represents the best candidate among the three eastern clouds, given the positional agreement with the hotspot. In \cite{tatematsu1990}, the density of cloud A is quoted to be of about 100 cm$^{-3}$. However, the impossibility to resolve the extension of hotspot, together with uncertainties on its distance, prevent us from assessing its total mass and therefore the interaction of the SNR with the northeastern cloud cannot be firmly established, as also argued in \cite{koo2001, byun2006} from CO observations.

Interestingly, in the southern part of the SNR, shocked molecular clumps with densities on the order of $10^2-10^4$\,cm$^{-3}$ have been observed \citep{koo2001}. This is so far the most convincing evidence for the interaction of the SNR with molecular clouds \citep{byun2006}. As previously reported, we find a hint of a gamma-ray excess in correspondence to this region. Whether the shocked molecular clouds are inside the SNR or lie along the same line of sight and appear located in the interior of the shell as a projection effect, they can be the sites of efficient gamma-ray production through enhanced $\pi^0$ decays. In the first scenario the gamma-ray emission could be explained in terms of trapped CRs \citep{Uchiyama:2010mu} and in the latter by escaping CRs \citep{aharonian1996emissivity, Gabici:2009ak}. In the context of shocked clouds inside the SNR, preexisting CRs might be shock-accelerated by the overrun of the blast waves and the reacceleration in the dense clouds might result in a hardening of the spectrum of existing particles. So far, the spectral properties of the southern excess cannot be solidly constrained owing to limited statistics. Therefore, more data are needed to highlight possible spectral variations with respect to the remnant.

\subsubsection{Bremsstrahlung} \label{cap:model_brem}
We hereafter consider models in which electron bremsstrahlung is dominant in the Fermi-LAT energy band. In order to reproduce the low energy drop of the gamma-ray emission, a break in the electron spectrum has to be postulated, as shown in Fig. \ref{fig:brem}. If a cutoff at lower energies is also included in the electron spectrum (i.e., when $E_{e,min} = 400$\,MeV), the Fermi data are satisfactorily well fitted in terms of bremsstrahlung emission. The shapes of the two components of the electron spectrum are estimated as $\alpha_{1,e} = 1.6 \pm 0.3$ and $\alpha_{2,e}= 3.1^{+0.2}_{-0.1} $, with a break around 2\,GeV. Such a spectrum notably evokes the spectrum of Galactic CR electrons in the interstellar medium (ISM), for which a low energy falloff with a break at $\sim2$\,GeV  followed by a second component with slope $\sim3.1$ are measured \citep{Strong:2015vrk, Strong:2011wd}. Remarkably, in this scenario the SNR would contribute to the acceleration of the diffuse component of the Galactic CR electrons.

In the leptonic scenario, we need to account for the radio emission of synchrotron origin. We assume the same parent spectral shape since in the easiest scenario the two emissions are produced by the same population of electrons, although the radio-shell morphology does not find a strict correspondence with the gamma-ray disk emission. In Fig. \ref{fig:hb21_radio} the radio data (taken from  \citep{kothes2006catalogue, Pivato:2013uoa}) are shown together with the synchrotron modeling performed with the Naima package \citep{naima}. Assuming a broken power-law spectrum and postulating $E_{e,min} = 400$\,MeV, the electron parameters are estimated as  $\alpha_{1,e} = 1.60 \pm 0.17$,  $\alpha_{2,e} =2.9^{+0.7}_{-0.4}$, $E_{b,e} = 6 \pm 3$\,GeV, and $B=25^{+6}_{-10}~\mu\mbox{G}$. The energy of the break is slightly higher than that found for the bremsstrahlung. To reproduce the observed radio break with an electron break at $E_{b,e} = 1.8$\,GeV (see Table \ref{tab:naima_results_brem} for the spectral parameters of the bremsstrahlung electrons), a magnetic field on the order of $\sim170 ~\mu\mbox{G}$ has to be postulated because of the $h\nu \propto E_{e}^2 B$ scaling of the characteristic energy of the synchrotron radiation. However, within the uncertainties, the synchrotron and bremsstrahlung energy breaks, as well as the other parameters of the electron spectrum, are in agreement. The missing correspondence between the radio the gamma-ray emission morphologies could eventually be explained in terms of different cooling times: in case of bremsstrahlung interactions the cooling timescale depends on the ambient density as $\tau_{brems} \sim 4 \times 10^7 (n_p /\mbox{ cm}^3)^{-1}$ years, while in case of synchrotron losses it scales as $\tau_{syn} \sim 1.3 \times 10 ^{10} (B / 1 \mu\mbox{G})^{-2} (E_e / 1 \mbox{ GeV})^{-1}$ years. The ratio of the two characteristic times cannot be significantly constrained owing to the uncertainties on the relevant quantities. It can vary from region to region of the SNR (e.g., due to the observed nonuniform gas distribution) and it is expected to be further amplified in the region of the shell because of the enhanced magnetic field. Higher resolution gamma-ray data might help to further investigate this scenario and potentially confirm the hypothesis of same production regions.
\begin{figure}[ht!]
\centering
\includegraphics[width=0.5\textwidth]{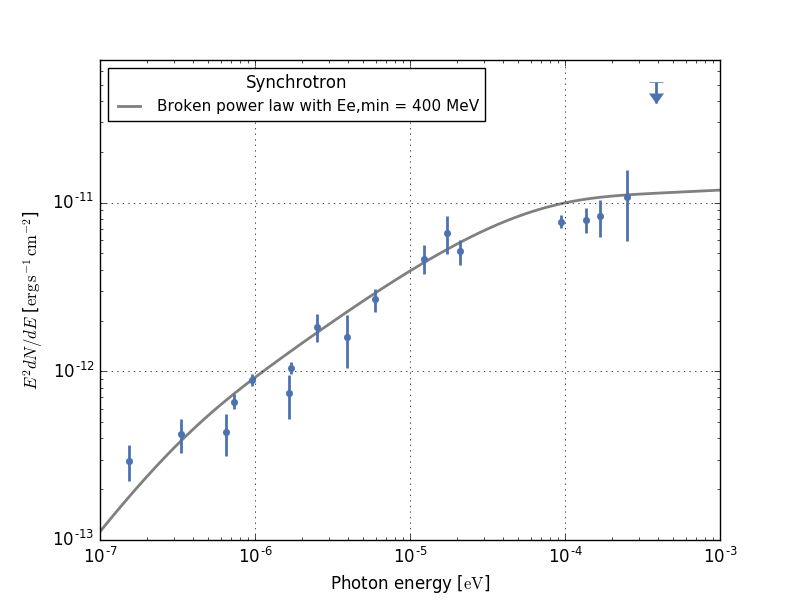}
\caption{\label{fig:hb21_radio}Modeling of the radio emission from HB 21. Radio data points are from the literature: up to $4\cdot10^{-5}$\,eV are from \cite{kothes2006catalogue}, above are WMAP data taken from \cite{Pivato:2013uoa}.}
\label{fig:hb21_radio}
\end{figure}

\subsection{Energetics}
The total energy content in protons required to explain the Fermi-LAT emission is on the order of $W_p \backsimeq 3 \times 10^{50} (n / 1 \mbox{ cm}^{-3})^{-1}$\,erg. A bremsstrahlung origin of the gamma rays implies for the parent electrons $W_e^{Brem} \backsimeq 2.1 \times 10^{49} (n / 1 \mbox{ cm}^{-3})^{-1}$\,erg. The energy budget in protons is high, although we cannot exclude a higher local gas density, which would reduce the required budget. For what concerns the electrons, from the modeling of the radio emission we obtained $W_e^{Synch} \backsimeq 1.7 \times 10^{48} (B / 25 ~\mu\mbox{G})^{-2}$\,erg. The comparison of $W_e^{Brem}$ and $W_e^{Synch}$ predicts a proton density of $\sim10\mbox{ cm}^{-3}$, which is reasonable for a SNR. We note that as long as the simple one-zone model is assumed and the same electrons produce both the radio and the gamma-ray photons, then the product $B^2 \times n^{-1}$ should be fixed. Consequently, for a magnetic field as high as $170 ~\mu\mbox{G}$, a density above hundreds of protons per cubic-centimeter is expected, which is unlikely within a SNR. In this sense, more precise radio measurements of the SNR are needed to better contain the position of the break, and therefore the magnetic field. This would eventually allow us to rule out the hypothesis of same parent electrons, together with a deeper understanding of the SNR environment.

\section{Conclusions}
\label{sec:conclusions}

We analyzed more than nine years of \emph{Fermi}-LAT Pass-8 data from the region of the SNR HB 21. The substantially richer statistics with respect to the previous analysis (based on roughly four years of Pass-7 data) led to a revision of the published results with relevant implications.
The source is centered at $l=88^{\circ}.76\pm 0^{\circ}.02, b=+4^{\circ}.74\pm0^{\circ}.02$ with an extension of $0^{\circ}.83 \pm 0^{\circ}.02$, that is, smaller than the extension previously published. The revised morphology allowed us to resolve a point-like source at the edge of the SNR, previously considered as part of HB 21. We proved that the bulk of the gamma-ray emission comes from the remnant and there is a hint of additional excess in correspondence to the region of the shocked molecular clouds observed at $4\sigma$. The spectral analysis confirmed the presence of a break in the gamma-ray emission. We demonstrated the significant detection of the break at the level of $5.4\sigma$ and an energy of $E_b = 389\pm95$\,MeV.

The discovery of a spectral break at low energies is crucial for the understanding of the origin of the radiation from the SNRs. We showed that the SED of the SNR HB 21 can be fitted by both leptonic and  hadronic gamma-ray production processes. A good agreement of the LAT data with the pion-decay mechanism is found, predicting a proton spectrum with a slope $\alpha\sim2.5$ and requiring a steepening at higher energies. The spectral and morphological properties of the high energy radiation from SNR HB 21 can be also interpreted as bremsstrahlung emission from a population of electrons accelerated within the SNR, whose spectrum resembles very closely the spectrum of CR electrons in the Galactic plane. This novel and remarkable result of our investigation shows that the SNR HB 21 contributes to the Galactic background electrons and, more in general, it can be treated as an indication of the major contribution of SNRs to relativistic electrons in the ISM. Within the uncertainties, the same population of electrons (with same spectral slopes and break) also explains the radio emission from the source. These results suggest that the break in the electron spectrum is not due to cooling processes, but rather reflects an intrinsic property of the source that is likely related to acceleration processes.

\begin{acknowledgements}
The authors would like to thank T. M. Dame and V. Zabalza for  useful discussions.  We thank the Max-Planck-Institut f\"ur Kernphysik as hosting institution. S. C. acknowledges the Polish Science Centre through grant agreement DEC-2017/27 / B / ST9 / 02272. R. Z. acknowledges the Alexander von Humboldt Foundation for financial support. 
\end{acknowledgements}

%
\bibliographystyle{aa} 
\bibliography{./biblio} 
%

\end{document}